**From *Animaculum* to single-molecules: 300 years of the light microscope**

Adam J. M. Wollman, Richard Nudd, Erik G. Hedlund, Mark C. Leake*

*Biological Physical Sciences Institute (BPSI), Departments of Physics and Biology, University of York, York YO10 5DD, UK.*

*mark.leake@york.ac.uk

**Abstract**

Although not laying claim to being the inventor of the light microscope, Antonj van Leeuwenhoek, (1632 –1723) was arguably the first person to bring this new technological wonder of the age properly to the attention of natural scientists interested in the study of living things (people we might now term 'biologists'). He was a Dutch draper with no formal scientific training. From using magnifying glasses to observe threads in cloth, he went on to develop over 500 simple single lens microscopes[1] with which he used to observe many different biological samples. He communicated his finding to the Royal Society in a series of letters[2] including the one republished in this edition of *Open Biology*. Our review here begins with the work of van Leeuwenhoek before summarising the key developments over the last ca. 300 years which has seen the light microscope evolve from a simple single lens device of van Leeuwenhoek's day into an instrument capable of observing the dynamics of single biological molecules inside living cells, and to tracking every cell nucleus in the development of whole embryos and plants.

1. **Antonj van Leeuwenhoek and invention of the microscope**

Prior to van Leeuwenhoek, lenses had existed for hundreds of years but it was not until the 17th century that their scientific potential was realized with the invention of the light microscope. The word 'microscope' was first coined by Giovanni Faber in 1625 to describe an instrument invented by Galileo in 1609. Gailieo's design was a compound microscope - it used an objective lens to collect light from a specimen and a second lens to magnify the image, but this was not the first microscope invented. In around 1590 Hans and Zacharias Janssen had created a microscope based on lenses in a tube[3]. No observations from these microscopes were published and it was not until Robert Hooke and Antonj van Leeuwenhoek that the microscope, as a scientific instrument, was born.

Robert Hooke was a contemporary of van Leeuwenhoek. He used a compound microscope, in some ways very similar to those used today with a stage, light source and three lenses. He made many observations which he published in his *Micrographia* in 1665.[4] These included seeds, plants, the eye of a fly and the structure of cork. He described the pores inside the cork as 'cells', the origin of the current use of the word in biology today.

Unlike Hooke, van Leeuwenhoek did not use compound optics but single lenses. Using only one lens dramatically reduced problems of optical aberration in lenses at the time, and in fact van Leeuwenhoek's instruments for this reason generated a superior quality of image to those of his contemporaries. His equipment was all handmade, from the spherical glass lenses to their bespoke fittings. His many microscopes consisted mainly of a solid base, to hold the single spherical lens in place, along with adjusting screws which were mounted and glued in place to adjust the sample holding pin, and sometimes an aperture placed before the sample to control illumination[1] (see Figure 1 for an illustration). These simple instruments could be held up to the sun or other light source such as a candle and did not themselves have any light sources inbuilt. His microscopes were very lightweight and portable, however, allowing them to be taken into the field to view samples as they were collected. Imaging consisted of often painstaking mounting of samples, focussing and then sketching, with sometimes intriguing levels of imagination, or documenting observations.

Van Leeuwenhoek's studies included the microbiology and microscopic structure of seeds, bones, skin, fish scales, oyster shell, tongue, the white matter upon the tongues of feverish persons, nerves, muscle fibres, fish circulatory system, insect eyes, parasitic worms, spider physiology, mite reproduction, sheep foetuses, aquatic plants and the 'animalcula' – the microorganisms described in his letter.[2] As he created the microscopes with the greatest magnification of his time he pioneered research into many areas of biology. He can arguably be credited with the discovery of protists, bacteria, cell vacuoles and spermatozoa.

2.     **The development of the microscope and its theoretical underpinnings**

It was not until the 19th century that the theoretical and technical underpinnings of the modern light microscope were developed. Most notably diffraction-limit theory, but also aberration-corrected lenses and an optimized illumination mode called Köhler illumination.

There is a fundamental limit to the resolving power of the standard light microscope; these operate by projecting an image of the sample a distance of several wavelengths of light from the sample itself, known as the 'far-field' regime. In this regime the diffraction of light becomes significant, for example through the circular aperture of the objective lens. This diffraction causes 'point sources' in the sample which scatter the light to become blurred spots when viewed through a microscope, with the level of blurring determined by the imaging properties of the microscope known as the point spread function (PSF). Through a circular aperture, such as those of lenses in a light microscope, the PSF can be described by a mathematical pattern called an Airy disk, which contains a central peak of light intensity surrounded by dimmer rings moving away from the centre (see Figure 2a). This phenomenon was first theoretically characterized by George Airy in 1835.[5] Later, Ernst Abbe would state that the limit on the size of the Airy disk was roughly half the wavelength of the imaging light[6] which agrees with the so-called Raleigh criterion for the optical resolution limit[7], which determines the minimum distance between resolvable objects (see Figure 2b). This became the canonical limit in microscopy for over a hundred years, with the only attempts to improve spatial resolution through the use of lower wavelength light or using electrons rather than photons, as in electron microscopy, which have a smaller effective wavelength by ~4 orders of magnitude.

Ernst Abbe also helped solve the problem of chromatic aberration. A normal lens focuses light to different points depending on its wavelength. In the 18th century, Chester Moore Hall invented the achromatic lens which used two lenses of different materials fused together to focus light of different wavelengths to the same point. In 1868 Abbe invented the apochromatic lens, using more fused lenses, which better corrected chromatic and spherical aberrations.[8] Abbe also created the world's first refractometer and we still use the 'Abbe Number' to quantify how diffraction varies with wavelength.[9] He also collaborated with Otto Schott, a glass chemist, to produce the first lenses which were engineered with sufficiently high quality to produce diffraction-limited microscopes.[10] Their work in 1883 set

the limits of far-field optics for over a century, until the advent of the 4π microscope in 1994.[11]

Another eponymous invention of Abbe was the Abbe condenser – a unit which focuses light with multiple lenses which improved sample illumination but was quickly superseded by Köhler Illumination, the modern standard for 'brightfield' light microscopy. August Köhler was a student of many fields of the 'natural sciences'. During his PhD studying limpet taxonomy he modified his illumination optics to include a field iris and also an aperture iris with a focusing lens to produce the best illumination with the lowest glare which aided in image collection using photosensitive chemicals.[12] Due to the slow nature of photography of the period good images required relatively long exposure times and Köhler Illumination greatly aided in producing high-quality images. He joined the Zeiss Optical Works in 1900, where his illumination technique coupled with the optics already developed by Abbe and Schott went on to form the basis of the modern brightfield light microscope.

### 3. Increasing optical contrast

One of the greatest challenges in imaging biological samples is their inherently low contrast, due to their refractive index being very close to water and thus generating little scatter interaction with incident light. A number of different methods for increasing contrast have been developed including imaging phase and polarization changes, staining and fluorescence, the latter being possibly the most far-reaching development since the invention of the light microscope.

Biological samples generate contrast in brightfield microscopy by scattering and absorbing some of the incident light. As they are almost transparent, the contrast is very poor. One way around this, is to generate contrast from phase (rather than amplitude) changes in the incident light wave. Fritz Zernike developed phase contrast microscopy in the 1930s[13] while working on diffraction gratings. Imaging these gratings with a telescope, they would 'disappear' when in focus.[14] These observations led him to realize the effects of phase in imaging, and their application to microscopy subsequently earned him the Nobel prize in 1953. Phase contrast is achieved by manipulating the transmitted, background light differently from the scattered light, which is typically phase-shifted 90 degrees by the sample. This scattered light contains information about the sample. A circular annulus is placed in front of the light source, producing a ring of illumination. A ring-shaped phase plate below the objective, shifts the phase of the background light by 90 degrees such that it is in phase (or sometimes completely out of phase, depending on the direction of the phase shift) with the scattered light producing a much higher contrast image.

An alternative to phase contrast is Differential Interference Contrast (DIC). It was created by Smith[15] and further developed by Georges Nomanski in 1955.[16] It makes use of a Normanski-Wollaston Prism through which polarized light is sheared into two beams polarized at 90 degrees to each other. These beams then pass through the sample and carry two brightfield images laterally displaced a distance equal to the offset of the two incoming beams at the sample plane. Both beams are focussed through the objective lens and then recombined through a second Normanski-Wollaston prism. The emergent beam goes through a final analyser emerging with a polarization of 135 degrees. The coaxial beams interfere with each other owing to the slightly different path lengths of the two beams at the same point in the image, giving rise to a phase difference and thus a high contrast image. The resultant image appears to have bright and dark spots which resemble an illuminated relief map. This faux relief map should not be interpreted as such, however, as the bright and dark spots contain information instead about path differences between the two sheared beams. The images produced are exceptionally sharp compared to other transmission modes. DIC is still the current standard technique for imaging unstained microbiological samples in having an exceptional ability to reveal the boundaries of cells and subcellular organelles.

Contrast can also be improved in biological samples by staining them with higher contrast material, for example dyes. This also allows differential contrast, where only specific parts of a sample, such as the cell nucleus, are stained. In 1858 came one of the earliest documented staining in microscopy when Joseph von Gerlach demonstrated differential staining of the nucleus and cytoplasm in human brain tissue soaked in the contrast agent carmine.[17] Other notable examples include silver staining introduced by Camillo Golgi in 1873, which allowed nervous tissue to be visualized, [18] and Gram staining invented by Hans Christian Gram in 1884,[19] which allowing differentiation of different types of bacteria. Sample staining is still widely in use today, including many medical diagnostic applications. However, the advent of fluorescent staining would revolutionize contrast enhancement in biological samples.

The word 'fluorescence' to describe emission of light at a different wavelength to the excitation wavelength was first made by Stokes in 1852.[20] Combining staining with fluorescence detection allows for enormous increases in contrast, with the first fluorescent stain fluorescein being developed in 1871.[21] In 1941, Albert Coons published the first work on Immunofluorescence. This technique uses fluorescently-labelled antibodies to label specific parts of a sample. Coons used a fluorescein derivative labelled antibody and showed that it could still bind to its antigen.[22] This opened the way to using fluorescent antibodies as a highly specific fluorescent stain.

Green fluorescent protein (GFP) was first isolated from the jellyfish, *Aequorea victoria*, in 1962[23] but it was not until 1994 that Chalfie et al.[24] showed that it could be expressed and fluoresce outside of the jellyfish. They incorporated it into the promoter for a gene that encoded β-tubulin and showed that it could serve as a marker for expression levels. The discovery and development of GFP by Osamu Shimomura, Martin Chalfie and Roger Tsien was recognised in 2008 by the Nobel prize in chemistry.

By mutating GFP, blue, cyan and yellow derivatives had been manufactured[25] but orange and red fluorescent proteins proved difficult to produce until the search for fluorescent proteins was expanded to non-bioluminescent organisms. This led to the isolation of dsRed from *Anthozoa,* a species of coral.[26] Brighter and more photostable fluorescent proteins were subsequently produced by directed evolution.[27] The discovery of spectrally distinct fluorescent proteins allowed multichannel (dual and multi-colour) fluorescence imaging and opened the way to studying the interaction between different fluorescently-labelled proteins.

Early work with fluorescent proteins simply co-expressed GFP on the same promoter as another gene to monitor expression levels. Proteins could also be chemically labelled outside of the cell and then inserted using microinjection.[25,28] A real breakthrough, with the discovery of GFP, was optimizing a method to fuse the genes of a protein of interest with a fluorescent protein and express this in a cell - thus leaving the cell relatively unperturbed. This was first demonstrated[29] on a GFP fusion to the *bcd* transcription factor in *Drosophila*.[30]

Fluorescent dyes have been used, not just high-contrast markers, but as part of molecular probes, which can readout dynamics between molecules and also environmental factors such as pH. In 1946, Theodore Förster posited that if a donor and acceptor molecule were sufficiently close together, non-radiative transfer of energy could occur between the two, now known as Förster resonance energy transfer (FRET), with efficiency proportional to the sixth power of the distance between them.[31] If such molecules are themselves fluorescent dyes then fluroescenbce can be used as a metric of putative molecular interaction through FRET. In 1967, Stryer and Haugland showed this phenomenon could be used as a molecular ruler over a length scale of ~1-10 nm.[32] Since then, FRET is used routinely to image molecular interactions and the distances between biological molecules, and also in fluorescence lifetime imaging (FLIM).[33] Fluorescent probes have also been developed to detect cell membrane voltages, local cellular viscosity levels and the concentration of specific ions, with calcium ion probes, for example, first introduced by Roger Tsien in 1980.[34]

## 4. The Fluorescence Microscope

The fluorescence microscope has its origins in ultraviolet (UV) microscopy. Abbe theory meant that better spatial resolution could be achieved using shorter wavelengths of light. August Köhler constructed the first UV microscope in 1904.[35] He found that his samples would also emit light under UV illumination (although he noted this as an annoyance). Not long after, Oskar Heimstaedt realized the potential for fluorescence and had a working instrument by 1911.[36] These transmission fluorescence microscopes were greatly improved in 1929 when Philipp Ellinger and August Hirt, placed the excitation and emission optics on the same side as the sample and invented the 'epifluorescence' microscope.[37] With the invention of dichroic mirrors in 1967[38], this design would become the standard in fluorescence microscopes. Several innovative illumination modes have also been developed for the fluorescence microscope, which have allowed it to image many different samples over a wide range of length scales. These modes include confocal, FRAP, TIRF, two-photon and light-sheet microscopy.

In conventional fluorescence microscopy, the whole sample is illuminated and emitted light collected. Much of the collected light is from parts of the sample which are out of focus. In confocal microscopy, a pinhole is placed after the light source such that only a small portion of the sample is illuminated and another pinhole placed before the detector such that only in-focus light is collected (see Figure 1). This can reduce the background in a fluorescence image and allow imaging further into a sample. The latter even enables optical sectioning and 3D reconstruction. The first confocal microscope was patented by Marvin Minsky in 1961.[39] This instrument preceded the laser so the incident light was not bright enough for fluorescence. With laser-scanning confocal microscopes[40] much better fluorescence contrast is achievable, as explored by White who compared the contrast in different human and animal cell lines.[41]

Fluorophores only emit light for a short time before they are irreversibly photobleached, and so microscopists must limit their sample's exposure to excitation light. Photobleaching can be used to reveal kinetic information about a sample by fluorescence recovery. In the earliest fluorescence recovery study, in 1974, Peters et al. bleached one half of fluorescein-labelled human erythrocyte plasma membranes and found that no fluorescence returned, indicating no observable mean diffusive process of the membrane over the experimental time scales employed.[42] Soon after, analytical work by Axelrod et al.[43] (on what they termed Fluorescence Photobleaching Recovery) allowed them to characterize different modes of diffusion in intracellular membrane trafficking. The term Fluorescence Recovery After Photobleaching (FRAP) appears to have been coined by Jacobson, Wu and Poste in 1976.[44] With FRAP capabilities commercially available on confocal systems, it is now widely used for measuring turnover kinetics in live cells.

When imaging features that are thin or peripheral such as cell membranes and molecules embedded in these, a widely used method is Total Internal Reflection Fluorescence (TIRF) microscopy. This technique uses a light beam introduced above the critical angle of the interface between the (normally) glass microscope coverslip and the water-based sample. The beam itself will be reflected by total internal reflection due to the differences in refractive index between the water and the glass, but at the interface an evanescent wave of excitation light is generated which penetrates only ~100 nm into the sample, thus only fluorophores close to the coverslip surface are excited, producing much higher signal-to-noise than conventional epifluorescence microscopy. It was first demonstrated on biological samples by Axelrod in 1981 to image membrane proteins in rat muscle cells and lipids in human skin cells.[45]

In conventional epifluorescence or even confocal, there is a limit to how far into the sample it is possible to image because of incident light scattering from the sample, creating a fluorescent background. This is particularly problematic when imaging tissues. Longer wavelength light scatters much less but few fluorophores can be excited by this with standard single photon excitation. In her doctoral thesis, in 1931, Maria Gopport-Mayer theorized that two photons with half the energy needed, can excite emission of one photon whose energy was the sum of the two photons during a narrow time window for absorption of ~$10^{-18}$ s.[46] The phenomenon of two-photon excitation (2PE) was not observed experimentally for another 30 years, until Kaiser and Garrett demonstrated it in CaF crystals.[47] The probability of 2PE occurring in a sample is low due to the very narrow time window of coincidence with respect to the two excitation photons, so high intensity light with a large photon flux is required to use the phenomenon in microscopy. In 1990 Denk used a laser in a confocal scanning microscope to image human kidney cells w 2PE.[48] Since then, it has become a powerful technique for observing molecular processes in live tissues, particularly in neuroscience, where the dynamics of neurons within a live rat brain were first observed by Svoboda et al.[49]

Another method of reducing background in fluorescent samples is to only illuminate the sample through the plane which is in focus. This can be achieved by shining a very flat excitation beam through the sample perpendicular to the optical axis. A. Voie et al. first demonstrated this, using light-sheet microscopy (LSM) in 1993.[50] LSM can be used to take fluorescence images through slices of a sample, allowing a stack of images to build a 3D reconstruction. One caveat of LSM is that samples need to be specially mounted to allow an unobstructed excitation beam as well as a perpendicular detection beam so a bespoke microscope is required. The technique was pioneered and developed by Ernst Stelzer in 2004, and termed Selective Plane Illumination Microscopy (SPIM), it was used to image live embryos in 3D.[51] Stelzer's group went on to image and track every nucleus in a developing zebrafish over 24 hours[52] and also the growth of plant roots at the cellular level in

*Arabidopsis.*[53] LSM has proven itself a powerful tool for developmental biology, the potential of which is only now being realized.

## 5.     Improving resolution in length and time

Fluorescence microscopy set new standards of contrast in biological samples that have enabled the technique to achieve possibly the ultimate goal of microscopy in biology and visualize single molecules in live cells. The Abbe diffraction limit, thought unbreakable for over one hundred years, has been circumvented by ever more inventive microscopy techniques which are now extending into three spatial dimensions.

The first single biological molecules detected were observed by Cecil Hall in the 1950s,[54] using electron microscopy of metallic fibres shadowed replicas of large, filamentous molecules including DNA and fibrous proteins, with dried samples in a vacuum. The very first detection of a single biological molecule in its functional *aqueous* phase was made by Boris Rotman, his seminal work published in 1961 involving the observation of fluorescently-labelled substrates of beta-galactosidase suspended in water droplets. The enzyme catalysed the hydrolysis of galactopyranose labelled with fluorescein to the sugar galactose plus free fluorescein, which had a much greater fluorescence intensity than when attached to the substrate. He could detect single molecules because each enzyme could turnover thousands of fluorescent substrate.[55] A more direct measurement was made by Thomas Hirschfield, in work published in 1976, who managed to see single molecules of globulin, labelled with ~100 fluorescein dyes, passing through a focused laser.[56] Single dye molecules were not observable directly until the advent of Scanning Near-field Optical Microscopy (SNOM) developed by Eric Betzig and Robert Chichester allowing them to image individual cyanine dye molecules in a sub-monolayer.[57] SNOM uses an evanescent wave from a laser incident on a ~100 nm probe aperture which illuminates a small section and penetrates only a small distance into the sample. Images are generated by scanning this probe over the sample. This is technically challenging as the probe must then be very close to the sample.

Single molecules were shown to be observable with less challenging methods when, using TIRF microscopy, single ATP turnover reactions in single myosin molecules was observed in

1995.[58] Other studies observed single F1-ATPase rotating using fluorescently labelled actin molecules in 1997[59] and the dynamics of single cholesterol oxidase molecules.[60] In a landmark study, the mechanism and step size of the myosin motor was determined by labelling one foot, observing and using precise Gaussian fitting to obtain nanometre resolution (termed 'fluorescence imaging with one nanometer accuracy' - FIONA).[61] This localization microscopy could effectively break the diffraction limit by using mathematical fitting algorithms to pinpoint the centre of a dye molecule's PSF image, as long as they are resolvable such that the typical nearest-neighbour separation of dye molecules in the sample is greater than the optical resolution limit. These techniques were soon applied to image single molecules in living cells[62,63] and now it is possible to count the number single molecules in complexes inside cells.[64,65]

Stefan Hell showed that it was possible to optically break the diffraction limit with a more deterministic technique which modified the actual shape of the PSF, called Stimulated Emission Depletion microscopy (STED), which he proposed with Jan Wichmann in 1994[66] and implemented with Thomas Klar in 1999.[67] STED works by depleting the population of excited energy state electrons through stimulated emission. Fluorescence emission only occurs subsequently from a narrow central beam inside the deactivation annulus region which is scanned over the sample. The emission region is smaller than the diffraction limit (~100 nm in the original study), thus allowing a superresolution image to be generated.

The development of STED showed that the diffraction limit could be broken and many new techniques followed. In 2002, Ando et al.[68] isolated a fluorescent protein from the stony coral, *Trachyphyllia geoffroyi,* which they named Kaede. They found that if exposed to UV light, its fluorescence would change from green to red and demonstrated this in Kaede protein expressed in HeLa cells. Photoactivatable proteins, such as this were used in 2006 by Hess et al. in Photo-Activated Localization Microscopy (PALM) using TIRF[69] and by Betzig et al. in Fluorescence Photo-Activated Localization Microscopy (FPALM) using confocal. Both methods use low intensity long UV laser light to photoactivate a small subset of sample fluorophores then another laser to excite them to emit and photobleach. This is repeated to build a superresolution image. A related method utilizes stochastic photoblinking of fluorescent dyes, which for example can be used to generate superresolution structures of DNA.[70]

Other notable superresolution techniques include Structured Illumination Microscopy (SIM).[71] In 1993, B. Bailey et al showed that structured stripes of light could be used to generate a spatial 'beat' pattern in the image which could be used to extract spatial features in the underlying sample image which had a resolution of ~2 times that of the optical resolution limit. In 2006, Xiaowei Zhuang *et al* demonstrated Stochastic Optical Reconstruction Microscopy (STORM)[72] which used a Cy5/Cy3 pair as a switchable probe. A red laser keeps Cy5 in a dark state and excites fluorescence, while a green laser brings the pair back into a fluorescent state. Thus, similarly to PALM, a superresolution image can be generated.

Improvements in *dynamic* fluorescence imaging have been significant over the past few decades. For example, using essentially the same localization algorithms as developed for PALM/STORM imaging, fluorescent dye tags can be tracked in a cellular sample in real-time for example tracking of membrane protein complexes in bacteria to nanoscale precicsion,[73] which have been extended into high time resolution dual-colour microscopy in vivo to monitor dynamic co-localization over with a spatial precision of ~10-100 nm.[74] Modifications to increase the laser excitation of several recent bespoke microscope systems have also improved the time resolution of fluorescence imaging down to the millisecond level, for example using narrowfield and Slimfield microscopy.[75]

3D information can be obtained in many ways including using interferometric methods[76] or multiplane microscopy[77] which image multiple focal planes simultaneously. Another method of encoding depth information in images is to distort the PSF image in an asymmetrical but measureable way as the light source moves away from the imaging plane. Astigmatism and double-helix microscopy accomplish this using different methods and are compatible with many modes of fluorescence illumination as the equipment used is placed between the objective lens and the camera. As such, it is a viable way to extract 3D data from many currently developed fluorescence microscopes.

Astigmatism microscopy is a simple 3D microscopy technique, first demonstrated by Kao and Verkman in 1994.[78] An asymmetry is introduced in the imaging path by placing a cylindrical lens before the camera detector. The introduced astigmatism offsets the focal plane along one lateral axis slightly, resulting in a controlled image distortion. When imaging singular or very small aggregates of fluorophores, the distortion takes the form of an ellipse, extending along either the *x* or *y* axis in the lateral plane of a camera detector conjugate to the microscope focal plane, depending on whether the fluorophore is above or below the focal plane. Values of 30 nm resolution in the lateral plane and 50 nm in the axial dimension have been reported using astigmatism with STORM.[79]

Double-helix PSF (DH-PSF) microscopy is a similar 3D microscopy technique using controlled PSF distortion. It exploits optical vortex beams, beams of light with angular momentum and works by placing a phase mask – an object which modifies the phase of the beam differently at different points along a cross-section – between the camera and the objective lens to turn the laser beam intensity profile from a Gaussian beam to a mixture of higher-order optical vortex beams - a superposition of two so-called Laguerre-Gauss (LG) beams. These two beams interfere with each other at the point that the light hits the camera creating two bright lobes.[80] The fields rotate as a function of distance propagated. As the two beams are superposed the distance is the same; if the two LG beams are slightly different the electric fields will rotate at different rates thanks to different so-called 'Gouy Phase' components. This means that the interference pattern produced rotates as a function of the distance of the point source from the image plane only.[81] The distance from the focal plane can be determined by measuring the rotation angle of the two lobes.

The phase mask can be created using transparent media such as etched glass or using a Spatial Light Modulator (SLM). An SLM is a 2D array of microscale bit components, each of which can be used to change the phase of the incident light across a beam profile. A liquid-crystal-on-silicon SLM retards light as a function of the input voltage to each bit. As such, a phase mask can be applied and changed in real-time using computer control. One major drawback is that they are sensitive to the polarization of light[82] limiting the efficiency of light propagation through the SLM. Alternatively a fixed glass phase plate can be etched using nanolithography. This is phase-independent and much more photon efficient. The phase is retarded simply by the thickness of the glass at each point in the beam. However, glass phase plates are less precise than SLM due to limitations in the lithography. Still, these are much easier to implement and can be purchased commercially or custom-built and used with almost any microscope setup with minimal detrimental impact. DH-PSF microscopy has been shown to have some of the smallest spatial localization errors of any 3D localization mode in high signal to noise systems.[83]

The power of beam-shaping combined with light-sheet illumination has been recently used to create lattice light-sheet microscopy.[84] Using Bessel beams, which focused laser profiles with minimal divergence due to diffraction, they create different bound optical lattices with different properties allowing them to image across four orders of magnitude in space and time and in diverse samples including diffusing transcription factors in stem cells, mitotic microtubules and embryogenesis in *Caenorhabditis elegans.*

## 6. The future

Although over 300 years since the pioneering work of van Leeuwenhoek, many of the major developments in light microscopy have occurred in just the past few decades and their full impact may not yet be felt. There are several technologies currently in development which may have a profound impact on microscopy. These include, for example, adaptive optics, lens-free microscopy, super lenses, miniaturization and combinational microscopy approaches.

A biological sample itself adds aberration through spatial variation in the refractive index. This is even more of a problem when imaging deep into tissues. Adaptive optics uses so-called dynamic correction elements such as deformable mirrors or SLMs to correct for this aberration, increasing spatial resolution and contrast. There have been many recent developments, reviewed comprehensively by Martin Booth,[85] but the technology is still yet to be widely adopted.

The archetypal lens used in light microscopy is made of glass, however this is not the only type of lens available. Optical diffraction gratings (optical gratings) can be used to focus, steer and even reflect light. Recognizing the need for miniaturization, researchers have been investigating the use of diffraction gratings in place of glass to help reduce the necessary size of optical components. While glass is great for large applications, it is extremely bulky when compared to the minimum size of a diffraction grating.[86] Optical gratings can be used as equivalent to lenses under some circumstances, for example a Fresnel Zone Plate can be used to focus light to a point as a convex lens does. Optical gratings all rely on the interaction of electromagnetic waves as they pass through the spaces in the gratings. This is fundamentally linked to the wavelength of propagating light making achromatic optical gratings very difficult to achieve in practice. Only recently have scientists been able to produce achromatic glass analogues such as an achromatic grating quarter-wave plates, for example, with good operational ranges.[87]

Ptychography completely removes the need for imaging optics, lenses or gratings, and directly reconstructs real-space images from diffraction patterns captured from a beam scanned over a sample. In many cases, this allows higher contrast images than DIC or phase contrast and 3D reconstruction.[88,89]

All optics currently used in microscopes are diffraction-limited but it is theoretically possible to construct, using so-called 'metamaterials', a perfect lens or super lens which could image with perfect sharpness. This was thought to require a material with negative refractive index[90] but it has now been shown that ordinary positive refractive index materials can also be used.[91] Even if super lenses are not achievable, new materials may revolutionize microscope lenses, still mostly composed of the same materials used by van Leeuwenhoek.

It is interesting to note the return of microscopes such as van Leeuwenhoek's which use only a single lens, in the foldascope developed by Manu Prakash at Stanford University[92] Using cardboard (an essential and surprisingly cheap component of some of the most advanced bespoke light microscopes found in our own laboratory) and simple filters and lenses, a near indestructible microscope with both normal transmission modes and fluorescence modes has been created that can be used by scientists and physicians working in areas far from expensive lab equipment.

Combinatorial microscopy is an interesting recent advance, which shows significant future potential. Here, several different microscopy methods are implemented on the same light microscope device. Many advances are being made at the level of single-molecule biophysics coupled to light microscopy in this regard. For example, methods being developed which can permit simultaneous superresolution imaging of DNA coupled to magnetic tweezers manipulation.[93]

The ultimate practical limits at the other end of the length scale for imaging tissues and whole organisms in the future are difficult to determine. Recent technological developments, such as the light-sheet imaging of *Arabidopsis* or lattice light-sheet microscopy discussed previously have enabled imaging of ever larger samples in greater detail. What limits the largest possible sample and to what level of detail it can be imaged is unknown. And, just as importantly, is computing technology used to store and analyse these data up to the challenge?

It is unquestionable that light microscopy has advanced enormously since the days of Antonj van Leeuwenhoek. The improvements have been, in a broad sense, twofold. Firstly, in length scale precision. This has been a 'middling-out' improvement, in that superresolution methods have allowed unprecedented access to nanoscale biological features, whereas light-sheet approaches and multi-photon deep imaging methods in particular have allowed incredible detail to be discerned at the much larger length scale level of multicellular tissues. Secondly, there has been an enormous advance, almost to the level of a paradigm shift, towards faster imaging in light microscopy, to permit truly dynamic biological processes to be investigated, right down to the millisecond level. Not only can we investigate detailed biological structures using light microscopy, but we can watch them change with time.

And yet, equally so, the basic principles of light microscopy for the study of biology remain essentially unchanged. These were facilitated in no small part by the genius and diligence of van Leeuwenhoek. It is perhaps the finest legacy for a true pioneer of light microscopy (.

**Author's contributions**

**Acknowledgements**

M.C.L. is supported by a Royal Society University Research Fellowship (UF110111).

E.G.H. was supported by Marie Curie EU FP7 ITN 'ISOLATE' ref 289995.

R.N. was supported by the White Rose Consortium.

The work was supported by the Biological Physics Sciences Institute (BPSI).

**Figures**

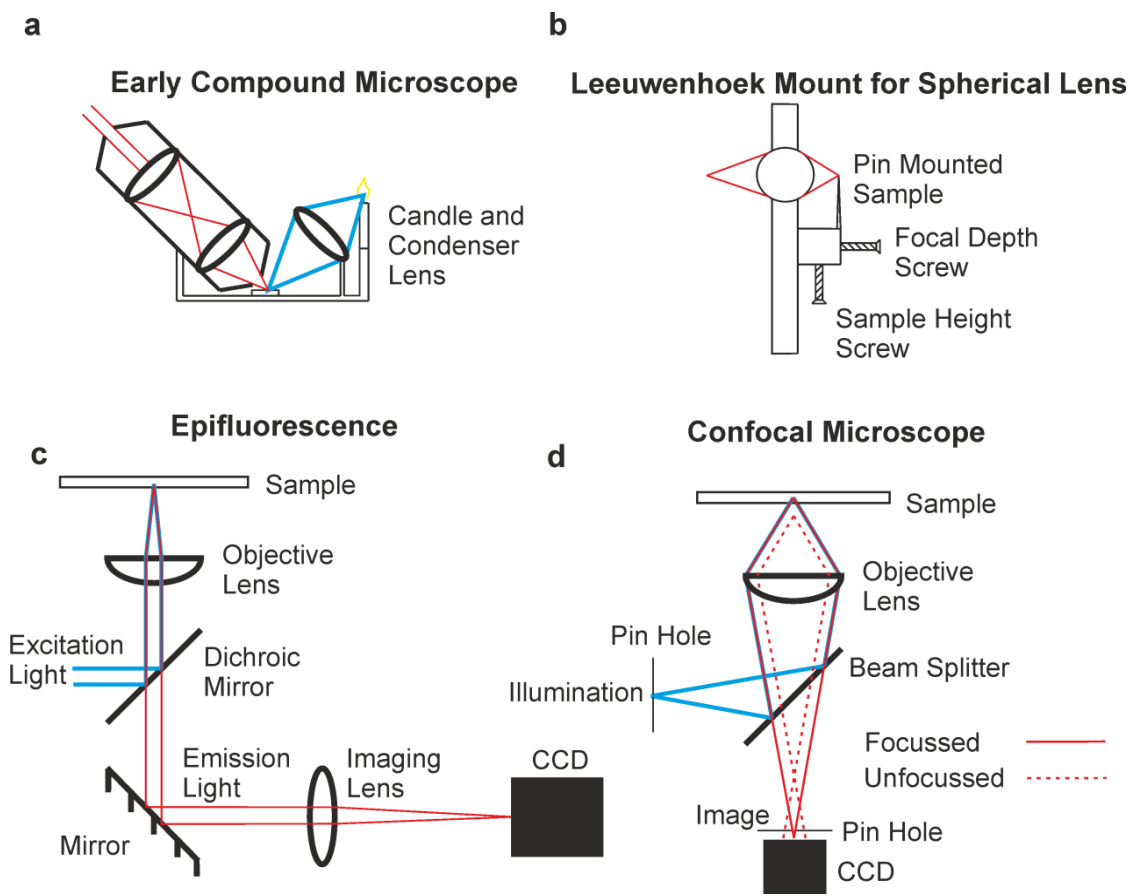

Figure 1

Optical microscope designs through the ages. a) One design of a simple compound microscope used by Hooke while writing Micrographia. b) An example of the single spherical lens mount system that van Leeuwenhoek used, approximately 5 cm in height. c) A simple epi-fluorescence system. d) A simple modern-day confocal microscope.

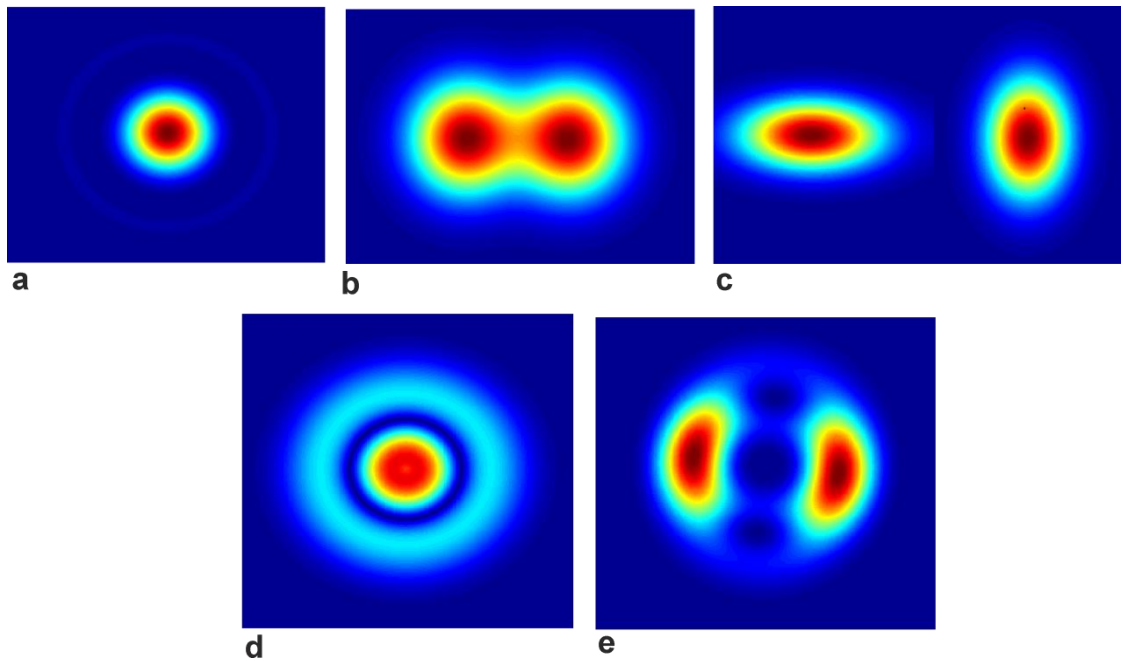

Figure 2

Mathematically generated Point Spread Function (PSF) images from in different light microscope designs. a) The Airy pattern, a disc and one of the rings produced by a point source emitter imaged using a spherical lens. b) Two such Airy discs separated by less than the Abbe limit for optical resolution. c) The lateral *xy* stretching exhibited in astigmatic imaging systems when the height *z* of a point source emitter is above or below the focal plane, the degree of stretching a metric for *z*. d) Expected pattern when a point source emitter is defocused. e) Two-lobed PSF used in Double-Helix PSF techniques, where the rotation of the lobes about the central point is used to calculate *z*.

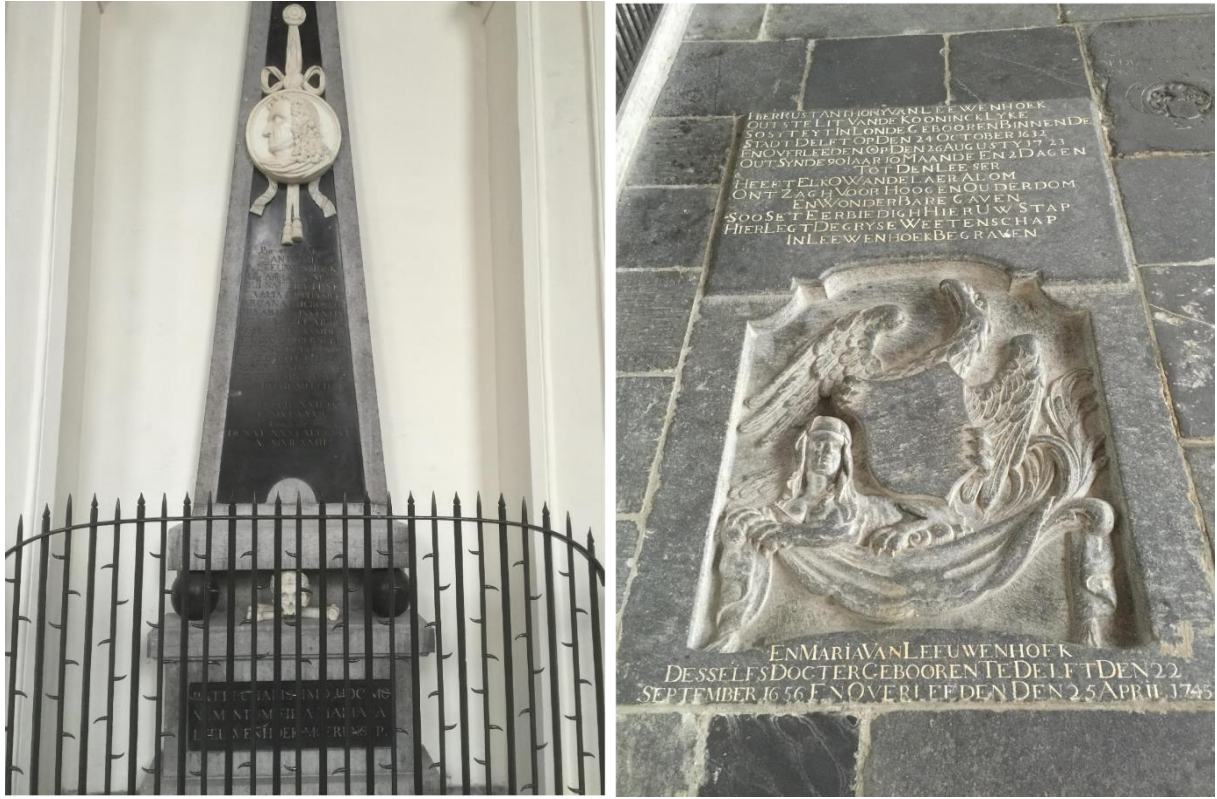

Figure 3

By chance, in the last days of finishing this review the corresponding author was staying ~100m from Leeuwenhoek's final resting place in the Oude Kerk, Delft, and captured these images.